\documentstyle[preprint,aps]{revtex}
\begin{document}%
\sloppy
\title{REVISITING BELL'S THEOREM FOR A CLASS OF DOWN-CONVERSION
EXPERIMENTS}
\author{ 
 \centerline{Sandu Popescu$^1$, Lucien Hardy$^2$ and Marek \.Zukowski$^3$}
\medskip
\centerline{\it $^1$Isaac Newton Institute, University of Cambridge,}
\centerline{\it Cambridge, CB3 0EH, U.K.}
\centerline{\it $^2$Dipartimento di Fisica, Universit\`a degli
Studi di Roma ``La Sapienza''}
\centerline{\it I-00185 Roma, Italy}
\centerline{\it $^3$Instytut Fizyki Teoretycznej 
i Astrofizyki, Uniwersytet Gda\'nski,}
\centerline{\it PL-80-952 Gda\'nsk, Poland}
}
\date{\today}
\maketitle
\thispagestyle{empty}\def \v {\vert}
\def \ra {\rangle}
\def \la {\langle}
\def \hx {\hat x}
\def \hy {\hat y}
\def \hz {\hat z}
\def \hxi {\hat \xi}
\def \he {\hat \eta}
\def \l {\lambda}

\begin{abstract}

A certain class of parametric down-conversion Bell type experiments has the
following features.
In the 
idealized perfect situation  
it is in only 50\% of cases that
each observer receives a
photon; in the other $50\%$ of cases one observer receives both
photons of a pair while the other observer receives none.
The standard approach is to discard the events of the second type.
Only the remaining ones are used as the data input to some 
Bell inequalities. This raises justified doubts whether such
experiments could be ever genuine tests of local realism. 
We propose, to take into account these
``unfavorable" cases and to
analyze the entire
pattern of polarization and localization correlations. 
This departure from the standard reasoning  
enables one to show that indeed the experiments 
are true test of local realism. 
\end{abstract}

The tremendous progress in experimental quantum optics during recent years,
and especially the advent of parametric down-conversion techniques have
generated a new wave of experiments purposing to test Bell's inequalities
[1].  The experiments like those of C. O. Alley and Y. H. Shih [2], Z. Y. Ou
and L. Mandel [3], and T. E. Kiss, Y. H. Shih, A. V. Sergienko and C. O.
Alley [4] are some of the most famous examples. They form a distinctive class:
polarization correlations are measured, and to produce the required phenomena
the use is made of a technique involving essentially two linear optical
devices, namely a wave plate and a single beamsplitter (fig.1).  
Although these experiments have successfully produced certain long-distance
quantum mechanical correlations, it is generally believed [5] that they could
never, {\it not even their idealized versions}, 
be considered as true tests of local realism. In the present
paper we prove that this general belief is wrong and that the above
experiments are in fact much better than they were previously thought to be:
in principle they {\it could} be tests of the most general premises of local
realism, and the only obstacles they face are purely technical (such as
present day low efficiency of the photon detectors, misalignments etc.).

The claim that the above experiments can never be true tests of local realism
is based usually on the following argument. In the original setting, proposed
by J. Bell and then widely analyzed by many others [6], experiments are
performed on an ensemble of pairs of particles prepared in such a way that
one particle in each pair is directed towards an observer while the other
particle is directed toward another observer situated far from the first one.
However, in the parametric down-conversion experiments we are interested in
[2-4] the pairs of particles (photons) are prepared in a different way - only
in $50\%$ of the cases each observer receives a photon; in the other $50\%$
of cases one observer receives both photons of a pair while the other
observer receives none.  The usual superficial method
to deal with this
situation is to discard all the unfavorable cases in
which both photons end at the same observer and to retain only the
cases in which each observer receives a photon. Therefore, 
it can be justifiably claimed, one runs directly into the well-known 
problem of
subensemble post-selection [6, 7]. If one restricts the analysis to a small 
enough
subensemble of the original ensemble of pairs one can never rule out all
possible local hidden variable models.

However, a more careful look at the the argument formulated in the last two
sentences shows its weakness. First of all, the idea that 
the ``unfavorable'' cases should be discarded appeared
probably because of the desire to forcefully map the problem under
investigation into the one originally formulated by Bell. But instead of
doing this one could actually make use of these ``unfavorable" cases by
taking them also into account and analyzing the entire pattern of
correlations. (This is the approach we shall take in the present
paper.) Second,
while it is widely believed that subensemble selection (when the selected
subensemble is small enough) {\it always} prevents observing violations of
local realism, this is actually not true. Indeed there are cases in which
post-selection is a serious drawback (e.g., ``detector efficiency" problem 
[7]) but there are also other
cases in which subensemble selection raises no problems [8, 9]. Consequently
one should not dismiss from the beginning the possibility of observing
violations of local realism just because subensemble selection has been
performed, but each situation should be investigated carefully.

Consider first the experimental setting used in [2-4] and illustrated in fig.
1. A type I parametric down-conversion source is used to generate pairs of
photons in which both photons have the same energy and linear polarization
(say $\hat x$) but propagate in two different directions. One of the photons
passes through a $90^0$ polarization rotator,
the wave-plate WP, emerging polarized along $\hat
y$. The two photons are then directed by two mirrors, M, 
onto the two sides of a (polarization independent) ``$50-50$''
beamsplitter BS which, for simplicity, we consider to be symmetric. Each
observer is equipped with a polarizing beamsplitter, orientated along an
arbitrary axis, randomly chosen just before the photons are supposed to
arrive. Each polarizing beamsplitter is followed by two detectors, $D_1^+$,
$D_1^-$ and $D_2^+$, $D_2^-$ respectively, where the lower index indicates
the corresponding observer and the upper index the two exit ports of the
polarized beamsplitter (``+" meaning parallel with the polarization axis of
the beamsplitter and ``-" meaning orthogonal to this axis). All
optical paths are assumed to be equal.

The quantum state of the two photons just before entering the detection
stations is $$\v\Psi\ra={1\over2}\bigl(\v 1\hx\ra_1\v 1\hy\ra_2-\v
1\hy\ra_1\v 1\hx\ra_2+i\v 1\hx,1\hy\ra_1\v 0\ra_2+i\v 0\ra_1\v
1\hx,1\hy\ra_2\bigr),\eqno(1)$$ where the subscript 1 or 2 on the ket vectors
represent the two regions of space where the photons arrive, i.e. near
observer 1 and near observer 2, the notation $0$ inside the ket vectors
denotes vacuum and $1\hx$ and $1\hy$ represents 1 photon polarized along the
$\hx$ or $\hy$ directions respectively.  The first two terms in (1)
correspond to cases in which each observer will register a single photon
while the last two terms correspond to cases in which one of the observers
will register two photons [10] while the other observer will register none. 

Quite often the discussion starts with just simply chopping off the
last two terms in (1). This
approach, since it is effectively a post-selection procedure, raises serious,
justified doubts [5], whether the experiments indeed are tests of local
realism. 
But in fact there is no reason why one should reject the other cases.

Let us denote by $P(i,\hxi; j,\he)$ the joint probability for the outcome
$i$  to be registered by observer 1 when his polarizing beamsplitter BS1 is
oriented along the direction $\hxi$ and the outcome $j$ to be registered by
observer 2 when her polarizing beamsplitter is oriented along $\he$. Here
$i,j=1-6$ and have the following meaning:

1 = one photon in $D^-$, no photon in $D^+$,

2 = one photon in $D^+$, no photon in $D^-$,

3 = no photons,

4 = one photon in $D^+$ and one photon in $D^-$,

5 = two photons in $D^+$, no photon in $D^-$,

6 = two photons in $D^-$, no photons in $D^+$.

We have not included the possibility of more than one pair of photons being
emitted. The probability of this happening, under the usual experimental
conditions, is very small, and thus will lead to  negligible effects.

A Clauser-Horne-Shimony-Holt (CHSH) type inequality which is obeyed by all
local hidden variables models but is violated by quantum mechanics for the
full state (1) can be obtained in the following way. Let us associate with
each outcome registered by observer 1 and 2 a corresponding value
$a_i^{\hxi}$ and $b_j^{\he}$ respectively, where $a_1^{\hxi}=b_1^{\he}=-1$
while all the other values are equal to 1, and let us denote by
$E(a^{\hxi}b^{\he})$ the expectation value of their product
$$E(a^{\hxi}b^{\he})=\sum_{i,j} a_i^{\hxi}b_j^{\he}P(i,\hxi;j,\he).\eqno(2)$$

Now, in a local hidden variables model $$P(i,\hxi;j,\he)=\int
d\l\rho(\l)P_1(i,\hxi; \l)P_2(j,\he; \l),\eqno(3)$$ where $\l$ is the local
hidden variable with $\rho$ its distribution function ($\int d\l\rho(\l)=1$),
and $P_1(i,\hxi; \l)$ and $P_2(j,\he; \l)$ the local probabilities ($\sum_i
P_1(i,\hxi; \l)=1=\sum_j P_2(j,\he; \l)$).  It is straightforward to see [6]
that
according to any local hidden variables model the CHSH inequality holds, i.e.
$$\v E(a^{\hxi}b^{\he})+ E(a^{\hxi}b^{\he'})+ E(a^{\hxi'}b^{\he})-
E(a^{\hxi'}b^{\he'})\v\le 2,\eqno(4)$$ for any directions $\hxi$, $\hxi'$,
$\he$ and $\he'$.

On the other hand, according to quantum mechanics $$
E(a^{\hxi}b^{\he})=\la\Psi\v A^{\hxi}B^{\he}\v \Psi\ra,\eqno(5)$$ where
$A^{\hxi}$ and $B^{\he}$ are the corresponding quantum observables, defined
as follows: $A^{\hxi}$ is an operator which has a nondegenerate eigenvalue
$A^{\hxi}=-1$ corresponding to the eigenstate $\v 1\hxi_{\perp}\ra_1$ (which
represents 1 photon polarized orthogonal to $\hxi$ and which corresponds to
observer 1 obtaining the outcome $i=1$) and a multiple degenerate eigenvalue
$A^{\hxi}=1$ corresponding to the rest of the Hilbert space (the space spanned
by $\v 1\hxi\ra_1$, $\v 0\ra_1$, $\v 1\hxi, 1\hxi_{\perp}\ra_1$, $\v
2\hxi\ra_1$ and $\v 2\hxi_{\perp}\ra_1$ and which correspond to observer 1
obtaining outcomes $i=2-6$); $B^{\he}$ is defined in a similar way. In other
words, the operators A and B are equal to the usual polarization operators on
the subspace of ``favorable" cases (yielding +1 if the photon's polarization
is parallel with that of the polarization analyzer and -1 if it is
perpendicular) and equal to the identity operator on the subspace of
``unfavorable" cases.  Let us also define $\v\Psi_1\ra$ and $\v\Psi_2\ra$ as
the normalized projections of $\v\Psi\ra$ on the subspaces of ``unfavorable"
and ``favorable" cases respectively, i.e.

$$\v\Psi_1\ra={i\over{\sqrt{2}}}\bigl(\v
1\hx,1\hy\ra_1\v 0\ra_2+\v 0\ra_1\v
1\hx,1\hy\ra_2\bigr)\eqno(6)$$
and 
$$\v \Psi_2\ra={1\over{\sqrt{2}}}\bigl(\v
1\hx\ra_1\v 1\hy\ra_2-\v 1\hy\ra_1\v 1\hx\ra_2\bigr)\eqno(7).$$

Then, as the local operators A and B do not mix the local vacuum and local
two photons states with the local one photon states, it follows that
$$CHSH_Q=\la \Psi\v
A^{\hxi}B^{\he}+A^{\hxi}B^{\he'}+A^{\hxi'}B^{\he}-A^{\hxi'}B^{\he'}\v\Psi\ra$$
$$=
{1\over2}\la \Psi_1\v
A^{\hxi}B^{\he}+A^{\hxi}B^{\he'}+A^{\hxi'}B^{\he}-A^{\hxi'}B^{\he'}\v\Psi_1\ra
+$$ $${1\over2}\la \Psi_2\v
A^{\hxi}B^{\he}+A^{\hxi}B^{\he'}+A^{\hxi'}B^{\he}-A^{\hxi'}B^{\he'}\v\Psi_2\ra
$$ $$={1\over2}\times2+{1\over2}\times\la \Psi_2\v
A^{\hxi}B^{\he}+A^{\hxi}B^{\he'}+A^{\hxi'}
B^{\he}-A^{\hxi'}B^{\he'}\v\Psi_2\ra\eqno(8)$$ The expectation value in the
last term in (8) is nothing else than the usual quantum CHSH expression
computed in the $\v\Psi_2\ra$ state which for suitable chosen directions
$\hxi$, $\hxi'$, $\he$ and $\he'$ can yield $2{\sqrt2}$.  Choosing such
directions in (8) it follows that, for the idealized perfect experiment 
$$CHSH_Q=1+\sqrt2>2\eqno(9)$$ which is in
contradiction with the limit imposed by local hidden variables models [11].

In calculating probabilities above we have assumed that the total number of
events is equal to the total number of pairs detected.  This is equivalent to
assuming an event ready configuration in which the source clicks (or gives
off some other appropriate signal) when the photons are emitted.  However,
the experiments [2-4] were not
event ready since there is no way to know that a pair of photons has
been emitted (event ready configurations have only been suggested [12]).

Thus, it could be possible, for example, 
that whether or not two photons are detected at
one end (whether we are in a i=4,5,6 or a i=3 case) depends on the setting of
the polarizing beamsplitter.  
Polarizer settings, by biasing the ensemble considered,
seemingly introduce the
possibility of a loophole.  

To solve this problem,  first of
all, we must decide what we are going to regard as an 
event. 
The experiment runs for a certain time, $T$.  This time can be divided up into
short intervals of duration $\tau$. 
At the beginning of each time interval the polarizing beamsplitters are set in 
a new, randomly chosen, direction. 
The time interval $\tau$ must be chosen to
be smaller than $L/c$ where $L$ is the distance between the
detector stations so that there is no possibility of relativistic causal
signals being transmitted during this time interval.  
Also, to avoid extra
complications, it should be chosen such that there is a very small
probability of more than one pair being emitted during $\tau$.  And it should
be bigger than the time resolution of the detectors so that two photons from
the same pair are almost certainly detected during the same time interval.  A
practical choice would probably be $\tau=10ns$.

The total number of events is $N$ where $T=N\tau$ and $T$ is chosen such that
$N$ is integer.  The $n$-th event happens during the interval $(n-1)\tau$ to
$n\tau$.  During this interval we record the type of event that has occurred
at each end of the apparatus ($i,j=1$ to $6$).  In this way we can form
probabilities in the usual way.  Typical counting rates are about $10^4$ per
second which is much smaller than $1\over\tau$.  This means almost all events
will be of the type where no photons are detected at either end ($i=j=3$).
For CHSH inequalities formulated in the usual way this would be a big problem
since these no-photon events would drown out the interesting events and hence
the inequalities would not be violated.  However, as we will see, the way in
which the correlation function has been defined in equation (2) solves this
problem.  

To include explicitly the vacuum term one can describe the state by 
$$
|\Psi'\rangle=\alpha|0\rangle_1|0\rangle_2 +\beta|\Psi\rangle \eqno(10)$$ 
where
$|\Psi\rangle$ is the state (1).  The vacuum term is now
explicitly included.  Either a total of two photons will be detected or no
photons.  Let $N_0$ be the number of events for which no photons are
detected.  Then $$ P(3,\hxi,3,\he)={N_0\over N}. \eqno(11) $$ Since
$a_3^{\hxi}=b_3^{\he}=1$ we have from equation (2) that $$
E(a^{\hxi}b^{\he})= {N_0\over N} + {N-N_0\over N}
\la\Psi\v A^{\hxi}B^{\he}\v \Psi\ra,\eqno(12)$$
Hence, now taking into account the vacuum cases, we have that
$$CHSH_Q={2N_0\over N} + {N-N_0 \over N} (1+\sqrt2)>2,\eqno(13)$$ where the
inequality follows since $N_0<N$. Hence, the CHSH inequalities are still
violated.  

The magnitude of the violation is not as great but this need not
bother us. Any experiment which violates 
the inequalities when the vacuum events are
included will also violate the experiment when they are not and vice versa.
Simply, an experiment where every event was a vacuum event would
give $CHSH_Q=2$.

Let us comment on the general case of non-event ready experiments.
In their review [6] Clauser and Shimony
make the point that the Clauser-Horne inequalities have a 0 as their
upper bound and hence they are insensitive to the overall normalization
of probabilities making them suitable to non-event ready experiments,
and since the CHSH inequalities have non-zero
bounds they do rely on knowing how to normalize probabilities and thus
are not suitable for non-event ready experiments.  However, by employing
the two tricks of (i) considering short intervals as events and (ii)
putting $a_3^{\hxi}=b_3^{\he}=1$ so that the CHSH inequality is saturated by
an ensemble of vacuum events, we make it possible to employ the CHSH
inequalities in non-event ready type experiments.

To summarize, if one wants to discuss
the experiments [2-4] as tests of local realism, 
one should not discard any ``unfavorable"
cases but rather one has to analyze them.  And
there is plenty of information we can obtain apart from what are the
polarization correlations. What we have just shown is that the entire
pattern of polarization and localization correlations in the 
experiments [2-4] cannot be explained by any local hidden
variable model.

The authors thank Paul Kwiat for his remark in [13] which 
in effect started the present 
collaboration by bringing our independent
works together. 
S.P. acknowledges the support of 
NSF Grant PHY-9321992. L.H. acknowledges CEE-TMR contract number
ERBFMRXCT96-0066.
M.\.Z. acknowledges the support of the University of 
Gda\'nsk research grants no. BW-5400-5-0087-6, and BW-5400-5-0306-7 and  
of the Batory Foundation.

\goodbreak
\bigskip
\noindent REFERENCES
\medskip
\nobreak
\small

\noindent
[1]  J. S. Bell, {\it Physics} {\bf 1} (1964) 195;
J. F. Clauser, M. A. Horne, A. Shimony, and R. A. Holt, {\it Phys. Rev.
Lett.} {\bf 23} (1969) 880.

\noindent
[2] C.O. Alley and  Y.H. Shih, Phys. Rev. Lett. {\bf 61} (1988) 2921. 

\noindent
[3] Z.Y. Ou and L. Mandel, Phys. Rev. Lett. {\bf 61} (1988) 50.

\noindent
[4]  T.E. Kiess, Y.H. Shih, A.V. Sergienko and C.O. Alley, Phys. Rev. Lett.
{\bf 71} (1993) 3893.

\noindent
[5] L. De Caro and A. Garuccio, Phys. Rev. A {\bf 50} (1994) R2803,
see also P.G. Kwiat, P.E. Eberhard, A.M. Steinberg and R.Y. Chiao, 
Phys. Rev. A {\bf 49}, 3209 (1994).  

\noindent
[6] J.F. Clauser and A. Shimony, Rep. Prog. Phys., {\bf 41}, 
1881 (1978).

\noindent
[7] D. Home and F. Selleri, Riv. N. Cim., {\bf 14}, 1 (1991).

\noindent
[8] S. Popescu, Phys. Rev. Lett.  72 (1994) 797;
S. Popescu, Phys. Rev. Lett. 74 (1995) 2619;
N. Gisin, Phys. Lett. A  151 (1996) 210;
A. Peres, 
Phys. Rev. A, {\bf 54}, 2685 (1996). 

\noindent
[9] B. Yurke and D. Stoler, Phys, Rev. A {\bf 46} (1992) 2229.

\noindent
[10] Even if one has at its disposal no detectors which can distinguish 
between
single counts and double counts, one can actually build such a detector out
of ordinary detectors (which cannot distinguish one- and two-photon events)
and beamsplitters. This can be realized by splitting the incoming beam into
$n$ beams by use of standard (non-polarizing) beam-splitters and placing an
ordinary detector in each of these $n$ beams. When the incoming beam
contains two photons, the probability that both photons end in just one of
the  $n$  beams  rapidly  goes  to  zero  as  $n$  increases.  Thus   for 
sufficiently
large $n$, almost always a  two-photon  incoming  state  will  result  in 
firing
of two of the detectors while a one-photon incoming state will fire just a
single detector. This may not be the case, if the detectors have a low 
efficiency. But also in the standard Bell type experiments one has very
high efficiency requiremnets, threfore this particular 
problem is not a specific feature 
of the studied case.

\noindent
[11] Instead of the CHSH inequality as used above,
one can equally well use the Clauser-Horne inequality. Obviously, in this case 
too,
information about the "unfavorable" cases has to be used.

\noindent
[12] M. \.Zukowski, A. Zeilinger, M.A. Horne and A.K. Ekert, Phys. Rev. Lett. 
{\bf 71}, 4287 (1993).

\noindent
[13] P.G. Kwiat, Phys. Rev. A {\bf52} 3380 (1995). 
The author proposes a different
state preparation method than the one of refs [2-3]. He also quotes 
our present idea (his footnote [9]).
 
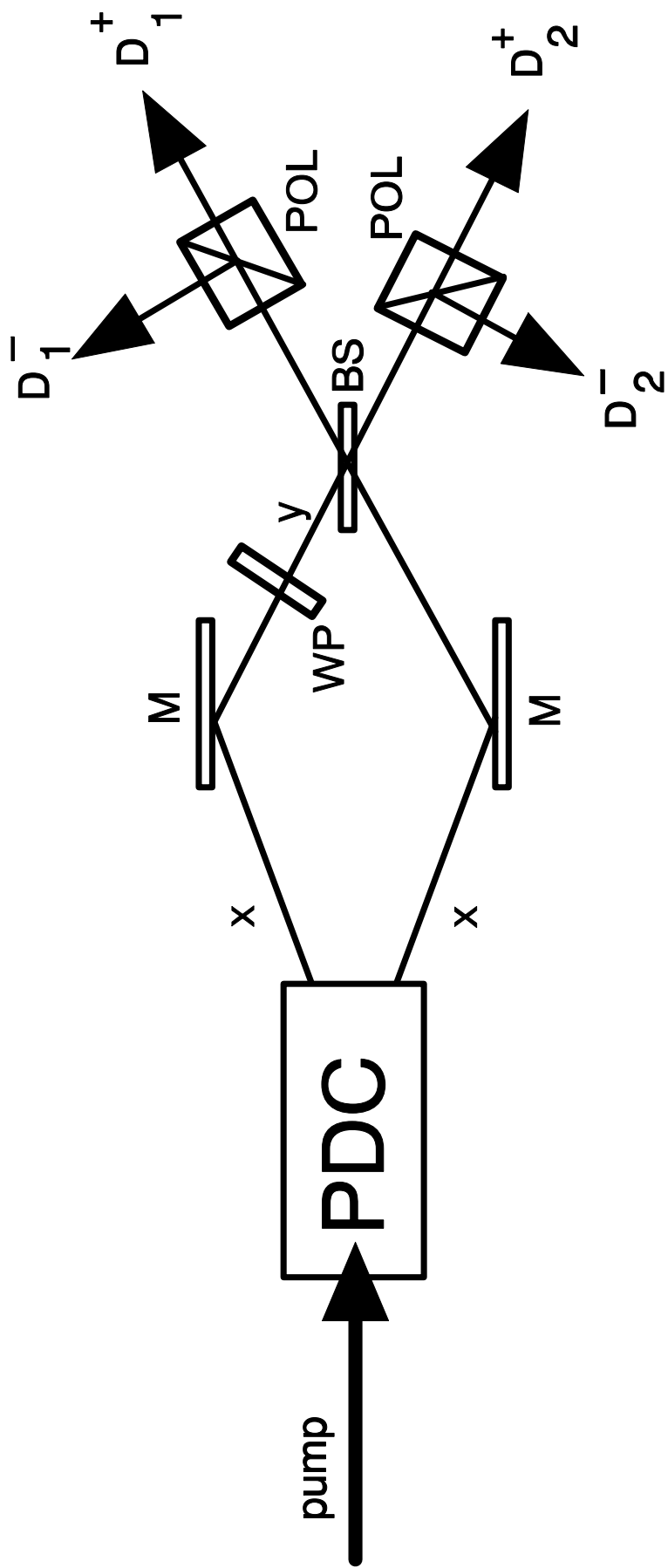
\end{document}